\documentclass[english]{article}
\usepackage[T1]{fontenc}
\usepackage[latin1]{inputenc}
\usepackage{color}

\makeatletter

\providecommand{\tabularnewline}{\\}

\usepackage{babel}
\makeatother
\begin{document}

\title{\textbf{\normalsize Parton Equilibration Enforcing Baryon Number
Conservation}\\
\textbf{\normalsize }\\
\textbf{\normalsize }\\
\textbf{\normalsize }\\
}

\author{{\normalsize Abhijit Sen}\\
{\normalsize Lecturer, Department of Physics, Suri Vidyasagar College,
Suri-731101, INDIA}\\
{\normalsize }\\
 {\normalsize }\\
{\normalsize }\\
{\normalsize }\\
 {\normalsize }\\
{\normalsize }\\
{\normalsize }
{\normalsize }\\
{\normalsize }\\
{\normalsize }\\
{\normalsize }\\
{\normalsize PACS No. 12.38.Mh(Quark Gluon Plasma), 21.65(Quark Matter),25.75(Relativistic
heavy ion collisions)}\\
{\normalsize }\\
{\normalsize }\\
{\normalsize }\\
{\normalsize }\\
}

\maketitle
\begin{abstract}
{\normalsize Parton equilibration for a thermally equilibrated but
chemically non-equilibrated quark-gluon plasma is presented. Parton
equilibration is studied enforcing baryon number conservation. Process
like quark - flavour interchanging is also considered. The degree
of equilibration is studied comparatively for the various reactions
/ constraints that are being considered. }\\
{\normalsize \par}
\end{abstract}
It is known {[} 1 ] that the initial system produced at RHIC energies
have a finite nonzero baryon number density. The anti particle to
particle ratio at mid rapidity for $\begin{array}{c}
\sqrt{s_{NN}}=130GeV\end{array}$ Au-Au collisions at the STAR collaboration,BNL report a noticiable
excess of baryons as compared to anti-baryons as reflected by the
yields of $\begin{array}{c}
\overline{p}\end{array}/p$ ( hovering between 0.6 and 0.8 ), $\begin{array}{c}
\overline{\Lambda}\end{array}/\Lambda$ ( hovering between 0.7 and 0.8 ) {[} 2 ] for varying transverse momentum,
centrality and rapidity. In a subsequent reporting of the BRAHMS Collaboration,
BNL {[} 3 ], it is seen that during Pb - Pb Collisions at $\begin{array}{c}
\begin{array}{c}
\sqrt{s_{NN}}=200GeV\end{array}\end{array}$, the $\begin{array}{c}
\overline{p}\end{array}/p$ ratio is a maximum of about 0.8 at zero rapidity and falls off for
higher rapidity values. 

This excess of baryons ( and hence quarks ) over anti-baryons ( and
hence anti-quarks ) clearly necessiates inclusion of a chemical potential
into the theoretical framework . This shall be the main point of emphasis
of the present piece of work. 

Inclusion of a chemical potential into the framework of parton equilibration
studies has been rather recent. Although in 1986, Matsui et. al. {[}
4 ] reported on strangeness equilibration rates at nonzero chemical
potential, it was not until the very end of the past century {[} 5,6
] that it was explicitly used in studying chemical equilibration processes.
A chemical potential for massless quarks that equalled the system
temperature at all temperatures was assumed by these authors. This
however was not very realistic as it did not take into account baryon
number conservation. Furthermore, they did not distinguish between
quarks and antiquarks. 

In 2004, He et. al. {[} 7 ] undertook a much complete study including
baryon number conservation. They used an expansion of the number densities
in powers of the chemical potential. They started with a quark- antiquark
distinction but that was soon put away. 

The present work is in the same line, but with the following points
of difference:\\
\\
1. We explicitly include the finite strange quark mass.\\
2. We maintain the quark-antiquark distinction.\\
3. We include the full phase space calculations for the pair production
and flavour changing processes in contrast with the factorised rates
of many earlier works. {[} 5,6,7,8,9,10 ]\\
4. For a given initial baryon number density we iterate to obtain
the initial value of the quark chemical potential which again evolves
obeying the baryon number conservation equation. \\
5. We study comparitively the degree of equilibration achieved for
varying initial conditions.\\

At this stage, let us note the following points as regard to the chemical
potential:\\
\\
1. Strangeness being very nearly conserved in strong interactions
and the initial strangeness content of the QGP fireball being zero
it implies that strange quark and antiquark are always produced in
pairs. This clearly indicates that the strange quark chemical potential
is zero.\\
2. Considering light quark-antiquark pair production processes we
can argue that since the net chemical potential on either side of
a chemical reaction ought to be the same and gluon chemical potential
is zero, this would necessarily imply \[
\mu_{q}=-\mu_{\overline{q}}\]
 \\
Again, since the mass of u and d quarks are nearly equal and much
less than that of the strange quark, we can safely treat the lighter
flavours in the same footing without any appreciable error being introduced.
By a suffix q we shall indicate the lighter quarks in general and
treat them as massless. \\
\\

\section{Basic Thermodynamics of the System}

The distribution functions for the constituent partons of the chemically
non-equilibrated state are taken to be as \\
${\displaystyle \begin{array}{c}
f_{g}\end{array}=\frac{\lambda_{g}}{e^{\varepsilon/T}-1}}$ .......... ( 1a )\\
$\begin{array}{c}
{\displaystyle f_{q(\overline{q})}=\frac{\lambda_{q(\overline{q})}e^{\pm\frac{\mu_{q}}{T}}}{e^{\varepsilon/T}+1}}\end{array}$.......... ( 1b ) and \\
$\begin{array}{c}
{\displaystyle f_{s}=f_{\overline{s}}=\frac{\lambda_{s}}{e^{\varepsilon/T}+1}}\end{array}$.......... ( 1c )\\
where the notations have usual meanings.

Of these, the first is the usual Bose distribution function {[} 8
], the second is the Modified Fermi-Dirac type distribution function
with an exponential term in the numerator to include the light quark
chemical potential {[}5 ] and the third is the usual modified Fermi-Dirac
distribution function for the massive strange quark {[} 8 ], each
being scaled by the non-equilibrium fugacity in order to describe
the degree of equilibrium achieved for a chemically equilibrating
system.

Using standard rules of Statistical Mechanics {[} 11 ] we can obtain
the number density $\begin{array}{c}
n\end{array}$, energy density$\begin{array}{c}
\varepsilon\end{array}$ and pressure $\begin{array}{c}
p\end{array}$ of the system. The results are generally represented as :\\
$\begin{array}{c}
{\displaystyle t=}\end{array}t_{g}+n_{f}(t_{q}+t_{\overline{q}})+2t_{s}$.......... ( 2 ) where $\begin{array}{c}
t=n,\varepsilon,p\end{array}$with\\
$\begin{array}{c}
{\displaystyle n=[\frac{16\zeta(3)}{\pi^{2}}\lambda_{g}+\frac{9\zeta(3)n_{f}}{2\pi^{2}}(\lambda_{q}e^{\frac{\mu_{q}}{T}}+\lambda_{\overline{q}}e^{-\frac{\mu_{q}}{T}})+}\end{array}$\\
$\begin{array}{c}
{\displaystyle \frac{6}{\pi^{2}}\lambda_{s}x_{s}^{3}\sum_{k=1}^{^{\infty}}(-1)^{k-1}\frac{K_{2}(kx_{s})}{(kx_{s})}]T^{3}}\end{array}$.......... ( 2a ) \\
$\begin{array}{c}
{\displaystyle \varepsilon=[\frac{8\pi^{2}}{15}\lambda_{g}+\frac{7\pi^{2}n_{f}}{40}(\lambda_{q}e^{\frac{\mu_{q}}{T}}+\lambda_{\overline{q}}e^{-\frac{\mu_{q}}{T}})+}\end{array}$\\
$\begin{array}{c}
{\displaystyle \frac{6}{\pi^{2}}\lambda_{s}x_{s}^{4}\sum_{k=1}^{^{\infty}}(-1)^{k-1}\{\frac{3K_{2}(kx_{s})}{(kx_{s})^{2}}+\frac{K_{1}(kx_{s})}{(kx_{s})}\}]T^{4}}\end{array}$.......... ( 2b ) \\
$\begin{array}{c}
\begin{array}{c}
{\displaystyle p=[\frac{8\pi^{2}}{45}\lambda_{g}+\frac{7\pi^{2}n_{f}}{120}(\lambda_{q}e^{\frac{\mu_{q}}{T}}+\lambda_{\overline{q}}e^{-\frac{\mu_{q}}{T}})+}\end{array}\end{array}$\\
$\begin{array}{c}
{\displaystyle \frac{6}{\pi^{2}}\lambda_{s}x_{s}^{4}\sum_{k=1}^{^{\infty}}(-1)^{k-1}\frac{K_{2}(kx_{s})}{(kx_{s})^{2}}]T^{4}}\end{array}$.......... ( 2c ) \\
where $\begin{array}{c}
n_{f}\end{array}$gives the number of massless quark flavours (=2 for our case) and
$\begin{array}{c}
x_{s}=\frac{m_{s}}{T}\end{array}$.

The baryon number density equals ( for two massless quark flavours)
\\
$\begin{array}{c}
{\displaystyle n_{b}=2X\frac{1}{3}(n_{q}-n_{\overline{q}})}\end{array}$.......... ( 3 ). For given initial values of baryon number density,
non-equilibrium fugacities and temperature we can iterate to obtain
the initial value of the chemical potential. The baryon number conservation
equation is given by $\begin{array}{c}
{\displaystyle \partial_{\mu}(n_{b}u^{\mu})=0}\end{array}$.......... ( 4 )\\
which may be expanded to obtain the rate of change of the chemical
potential with time. We obtain \\
\\
$\dot{\mu_{q}}=\dot{\lambda_{q}}B_{1}+\dot{\lambda_{\overline{q}}}B_{2}+\dot{T}B_{3}+B_{4}$
......... (5 )where \\
$\begin{array}{c}
{\displaystyle B_{1}=\frac{-Te^{\mu_{q}/T}}{\lambda_{q}e^{\mu_{q}/T}+\lambda_{\overline{q}}e^{-\mu_{q}/T}}}\end{array}$......... ( 5a )\\
$\begin{array}{c}
{\displaystyle B_{2}=\frac{Te^{-\mu_{q}/T}}{\lambda_{q}e^{\mu_{q}/T}+\lambda_{\overline{q}}e^{-\mu_{q}/T}}}\end{array}$......... ( 5b )\\
$\begin{array}{c}
{\displaystyle B_{3}=\frac{\mu_{q}}{T}-3\frac{\lambda_{q}e^{\mu_{q}/T}-\lambda_{\overline{q}}e^{-\mu_{q}/T}}{\lambda_{q}e^{\mu_{q}/T}+\lambda_{\overline{q}}e^{-\mu_{q}/T}}}\end{array}$ ......... ( 5c ) and \\
$\begin{array}{c}
{\displaystyle \begin{array}{c}
{\displaystyle B_{4}=-\frac{T(\lambda_{q}e^{\mu_{q}/T}-\lambda_{\overline{q}}e^{-\mu_{q}/T})}{\tau(\lambda_{q}e^{\mu_{q}/T}+\lambda_{\overline{q}}e^{-\mu_{q}/T})}}\end{array}}\end{array}$......... (5d )\\
It is to be noted here that for a baryon-free plasma $\dot{\begin{array}{c}
\mu_{q}\end{array}}$ vanishes. As we shall see shortly, due to these non-zero coefficients,
the light quark and anti-quark number density evolution equations
get coupled with one another.

\section{Parton Equilibration Equations}

The fundamental equation that dictates the parton number density evolution
is given by \\
\[
{\displaystyle \partial_{\mu}}(n_{k}u^{\mu})=\frac{\partial n_{k}}{\partial\tau}+\frac{n_{k}}{\tau}=(R_{gain}-R_{loss})\]
$\qquad\qquad\qquad\qquad\qquad\qquad\qquad\qquad\qquad\qquad\qquad$........
(6 )\\
where the RHS gives the difference of rate of gain and loss of the
parton species k i.e the net rate of change of the number density.
Using standard procedures {[} 10, 11 ] this equation leads to the
parton number density evolution equations as we shall see shortly.
\\

\subsection{{\normalsize Gluon Number Density Evolution Equation}\protect \\
}

The Gluon Number Density Evolution Equation is given as \\
\[
{\displaystyle \partial_{\mu}}(n_{g}u^{\mu})=(R_{gg\rightarrow ggg}-R_{ggg\rightarrow gg})-\sum_{i}(R_{gg\rightarrow i\overline{i}}-R_{i\overline{i}\rightarrow gg})\]
$\qquad\qquad\qquad\qquad\qquad\qquad\qquad\qquad\qquad\quad\qquad\qquad$........
(7 )\\
where the sum is over all quark flavours present. Substituting for
the gluon number density and substituting for the rates we have \[
{\displaystyle \dot{\lambda_{g}}G_{1}+}\dot{T}G_{2}+G_{3}=0\]
$\qquad\qquad\qquad\qquad\qquad\qquad\qquad\qquad\qquad\quad\qquad\qquad$........
(8 )\\
where\\
\\
$G_{1}=1/\lambda_{g}$........ (8a )\\
$\begin{array}{c}
G_{2}=3/T\end{array}$........ (8b ) and \\
$\begin{array}{c}
G_{3}=(1/\tau)-(R_{3}(1-\lambda_{g})-\sum_{i}R_{2i}/n_{g}\end{array})$........ (8c )\\
where, as in earlier works {[} 8, 10, 11 ] we have introduced rates
$\begin{array}{c}
R_{2i}\end{array}$and $\begin{array}{c}
R_{3}\end{array}$. We shall look more closely at these rates soon.\\

\subsection{{\normalsize Quark and Anti-Quark Number Density Evolution Equations}\protect \\
{\normalsize }\protect \\
}

As mentioned earlier, due to the coefficients $\begin{array}{c}
{\displaystyle B_{i}}\end{array}$, the light quark and anti-quark number density evolution equations
get coupled to each other. We obtain the following results:\\
\\

\subsubsection{Massless Quarks\protect \\
\protect \\
}

The equation is given by \\
\\
${\displaystyle \dot{\lambda_{q}}Q_{1}{\displaystyle +}\dot{\lambda_{\overline{q}}}Q_{2}+\dot{T}Q_{3}+Q_{4}=0}$........
(9 )\\
where\\
$\begin{array}{c}
{\displaystyle Q_{1}=\frac{1}{\lambda_{q}}+\frac{B_{1}}{T}}\end{array}$........ ( 9a )\\
$\begin{array}{c}
\begin{array}{c}
{\displaystyle Q_{2}=\frac{B_{2}}{T}}\end{array}\end{array}$........ (9b )\\
$\begin{array}{c}
{\displaystyle Q_{3}=\frac{3}{T}}\end{array}{\displaystyle +\frac{B_{3}}{T}-\frac{\mu_{q}}{T^{2}}}$........ (9c )\\
${\displaystyle \begin{array}{c}
{\displaystyle Q_{4}=}{\displaystyle \frac{1}{\tau}+\frac{B_{4}}{T}-SQ}\end{array}}$........ (9d )with\\
$\begin{array}{c}
SQ=\end{array}\{(R_{gg\rightarrow q\overline{q}}-R_{q\overline{q}\rightarrow gg})/n_{q}\}-\{(R_{q\overline{q}\rightarrow s\overline{s}}-R_{s\overline{s}\rightarrow q\overline{q}})/n_{q}\}$........ (9e ) \\
for light quarks \\
\\

\subsubsection{Massless Anti-Quarks\protect \\
\protect \\
}

For the anti-quark case, we obtain the equation as\\
\\
$\begin{array}{c}
{\displaystyle \dot{\lambda_{q}A}Q_{1}{\displaystyle +}\dot{\lambda_{\overline{q}}}AQ_{2}+\dot{TA}Q_{3}+AQ_{4}=0}\end{array}$ ........ ( 10 )where 

${\displaystyle \begin{array}{c}
{\displaystyle {\displaystyle AQ_{1}=-\frac{B_{1}}{T}}}\end{array}}$........ ( 10a )\\
${\displaystyle \begin{array}{c}
{\displaystyle \begin{array}{c}
{\displaystyle AQ_{2}=\frac{1}{\lambda_{\overline{q}}}-\frac{B_{2}}{T}}\end{array}}\end{array}}$........ ( 10b )\\
$\begin{array}{c}
{\displaystyle \begin{array}{c}
{\displaystyle AQ_{3}=\frac{3}{T}}\end{array}{\displaystyle -\frac{B_{3}}{T}+\frac{\mu_{q}}{T^{2}}}}\end{array}$........ ( 10c )\\
$\begin{array}{c}
{\displaystyle {\displaystyle \begin{array}{c}
{\displaystyle AQ_{4}=}{\displaystyle \frac{1}{\tau}-\frac{B_{4}}{T}-SaQ}\end{array}}}\end{array}$........ ( 10d )with \\
$\begin{array}{c}
\begin{array}{c}
SaQ=\end{array}\{(R_{gg\rightarrow q\overline{q}}-R_{q\overline{q}\rightarrow gg})/n_{\overline{q}}\}-\{(R_{q\overline{q}\rightarrow s\overline{s}}-R_{s\overline{s}\rightarrow q\overline{q}})/n_{\overline{q}}\}\end{array}$........ ( 10e )\\
for light anti-quarks\\
\\

\subsubsection{Massive Strange Quarks\protect \\
\protect \\
}

For massive strange quark the equation is given by \\
\\
${\displaystyle \dot{\lambda_{s}}S_{1}+\dot{T}S_{2}+S_{3}=0}$........
( 11 )\\
where\\
$\begin{array}{c}
{\displaystyle S_{1}=\frac{1}{\lambda_{s}}}\end{array}$........ ( 11a ) \\
${\displaystyle S_{2}=}{\displaystyle \frac{3}{T}}\frac{{\displaystyle \sum_{k=1}^{^{\infty}}(-1)^{k-1}\{\frac{3K_{2}(kx_{s})}{(kx_{s})^{2}}+\frac{K_{1}(kx_{s})}{(kx_{s})}\}}}{{\displaystyle \sum_{k=1}^{^{\infty}}(-1)^{k-1}\frac{K_{2}(kx_{s})}{(kx_{s})}}}$........
( 11b ) \\
${\displaystyle \begin{array}{c}
{\displaystyle S_{3}=}{\displaystyle \frac{1}{\tau}-SQs}\end{array}}$........ ( 11c ) with \\
$\begin{array}{c}
\begin{array}{c}
SQs=\end{array}\{(R_{gg\rightarrow s\overline{s}}-R_{s\overline{s}\rightarrow gg})/n_{s}\}+2\{(R_{q\overline{q}\rightarrow s\overline{s}}-R_{s\overline{s}\rightarrow q\overline{q}})/n_{s}\}\end{array}$........ ( 11d ) \\
for strange quarks\\

\subsection{{\normalsize Energy-Momentum Conservation Equation}\protect \\
{\normalsize }\protect \\
}

From the energy- momentum conservation equation $\begin{array}{c}
\partial_{\mu}T^{\mu\nu}=0\end{array}$........ ( 12 ) \\
 we obtain for (1+1)D $\begin{array}{c}
{\displaystyle {\displaystyle \frac{d\varepsilon}{d\tau}+\frac{\varepsilon+p}{\tau}}=0}\end{array}$........ ( 13 ) \\
( which we shall refer to as Bjorken's equation henceforth). We get
on substituting for energy and momentum\\
$\begin{array}{c}
{\displaystyle \begin{array}{c}
\dot{T}f_{6}+\dot{\lambda_{g}}f_{7}+\dot{\lambda_{q}}f_{8}+\dot{\lambda_{\overline{q}}}f_{9}+\dot{\lambda_{s}}f_{s}+f_{10}=0\end{array}}\end{array}$........ ( 14 )\\
where\\
\\

${\displaystyle \begin{array}{c}
{\displaystyle f_{6}=\frac{32\pi^{2}}{15}\lambda_{g}T^{3}+\frac{{\displaystyle 7{\displaystyle \pi^{2}}}}{20}(\lambda_{q}e^{\frac{\mu_{i}}{T}}-\lambda_{\overline{q}}e^{-\frac{\mu_{i}}{T}}){\displaystyle (\frac{B_{3}}{T}-\frac{\mu_{q}}{T^{2}})T^{4}}+\frac{{\displaystyle 28{\displaystyle \pi^{2}}}}{20}(\lambda_{q}e^{\frac{\mu_{i}}{T}}+\lambda_{\overline{q}}e^{-\frac{\mu_{i}}{T}}){\displaystyle T^{3}}+}\end{array}}$\\
$\begin{array}{c}
{\displaystyle {\displaystyle \frac{72}{\pi^{2}T}m_{s}^{4}\lambda_{s}}\{\frac{K_{2}(kx_{s})}{(kx_{s})^{2}}}{\displaystyle +\frac{5K_{1}(kx_{s})}{12(kx_{s})}+\frac{K_{0}(kx_{s})}{12}\}}\end{array}$........ ( 14 a)\\
\\
$\begin{array}{c}
{\displaystyle f_{7}=\frac{8\pi^{2}}{15}T^{4}}\end{array}$........ ( 14b )\\
$\begin{array}{c}
f_{s}=\frac{6}{\pi^{2}}m_{s}^{4}\sum_{k=1}^{^{\infty}}(-1)^{k-1}\{\frac{3K_{2}(kx_{s})}{(kx_{s})^{2}}+\frac{K_{1}(kx_{s})}{(kx_{s})}\}\end{array}$........ ( 14c)\\
$\begin{array}{c}
{\displaystyle f_{8}=\begin{array}{c}
{\displaystyle \frac{7\pi^{2}}{20}e^{\frac{\mu_{q}}{T}}T^{4}}\end{array}+\frac{7\pi^{2}}{20}(\lambda_{q}e^{\frac{\mu_{q}}{T}}-\lambda_{\overline{q}}e^{-\frac{\mu_{q}}{T}})}\frac{B_{1}}{T}T^{4}\end{array}$........ ( 14d )\\
$\begin{array}{c}
{\displaystyle \begin{array}{c}
{\displaystyle f_{9}=\begin{array}{c}
{\displaystyle \frac{7\pi^{2}}{20}e^{-\frac{\mu_{q}}{T}}T^{4}}\end{array}+\frac{7\pi^{2}}{20}(\lambda_{q}e^{\frac{\mu_{q}}{T}}-\lambda_{\overline{q}}e^{-\frac{\mu_{q}}{T}})}\frac{B_{2}}{T}T^{4}\end{array}}\end{array}$........ ( 14e )\\
$\begin{array}{c}
{\displaystyle f_{10}=\frac{32\pi^{2}}{45\tau}\lambda_{g}T^{4}+\frac{{\displaystyle 28{\displaystyle \pi^{2}}}}{60\tau}(\lambda_{q}e^{\frac{\mu_{i}}{T}}+\lambda_{\overline{q}}e^{-\frac{\mu_{i}}{T}}){\displaystyle T^{4}}+}\end{array}$\\
$\begin{array}{c}
{\displaystyle \frac{6m_{s}^{4}}{\pi^{2}\tau}\lambda_{s}\Sigma_{k=1}^{\infty}(-1)^{k-1}\{\frac{4K_{2}(kx_{i})}{(kx_{i})^{2}}+\frac{K_{1}(kx_{i})}{(kx_{i})}\}}\end{array}$\\
$\begin{array}{c}
{\displaystyle +\frac{7\pi^{2}}{20}(\lambda_{q}e^{\frac{\mu_{i}}{T}}-\lambda_{\overline{q}}e^{-\frac{\mu_{i}}{T}})\frac{B_{4}}{T}T^{4}}\end{array}$........ ( 14f )\\
\\
There remains a very important point to mention regarding the evaluation
of the rate of change of the chemical potential. Since the baryon
number conservation equation is obtained directly from the quark and
antiquark number densities, as are the massless quark and anti-quark
number density evolution equations, these two evolution equations
(coupled by the four B - coefficients) alongwithwith the energy-momentum
conservation equation ( with substitutions for $\dot{\begin{array}{c}
\lambda_{s}\end{array}}$, $\begin{array}{c}
\dot{\begin{array}{c}
\lambda_{g}\end{array}}\end{array}$ from the respective number density evolution equations) do not generate
a set of solvable coupled equations for $\begin{array}{c}
{\displaystyle \dot{\begin{array}{c}
\lambda_{q}\end{array}}}\end{array}$, $\begin{array}{c}
{\displaystyle \begin{array}{c}
{\displaystyle \dot{\begin{array}{c}
\lambda_{\overline{q}}\end{array}}}\end{array}}\end{array}$,$\begin{array}{c}
{\displaystyle \dot{T}}\end{array}$ as the system deeterminant vanishes.

To tackle this problem, we adopt an approximation scheme. While evaluating
$\begin{array}{c}
{\displaystyle B_{j},(j=1,2)}\end{array}$we use a truncated form of the expansion of the exponential in the
numerator keeping the rest of the expressions unchanged.

\section{Parton Equilibration Rates}

\subsection{{\normalsize Gluon Multiplication Rate $\protect\begin{array}{c}
R_{3}\protect\end{array}$}}

The gluon multiplication rate has been calculated by Xiong et. al.
{[} 12 ]. By explicitly calculating the matrix element {[} 13 ] (summed
over all the final states and averaged over all initial states) we
can obtain the gluon multiplication rate. \\
However, to avoid the huge calculations of evaluating 25 Feynman diagrams
{[} 13 ] involved, we fall back on the treatments used in earlier
works {[} 6,11 ]. We postulate that the gluon multiplication rate
depends on the chemical potential via the Debye Screening mass.\\
The Debye Screening mass suitable for a multicomponent chemically
non-equilibrated parton plasma is given by {[} 14 ]\\
$\begin{array}{c}
{\displaystyle m_{D}^{2}=\frac{2g^{2}}{\pi^{2}}\int dk}k[N_{c}f_{g}+\sum_{i}f_{i}]\end{array}$........ ( 15 )\\
where the sum runs over all flavours i, while $\begin{array}{c}
N_{c}\end{array}$gives the number of colours. To accomodate for antiquarks and remembering
that our number of flavours is 3 and not 6, we propose the following
modification:\\
$\begin{array}{c}
{\displaystyle \begin{array}{c}
{\displaystyle m_{D}^{2}=\frac{2g^{2}}{\pi^{2}}\int dk}k[3f_{g}+\begin{array}{c}
{\displaystyle \frac{1}{2}}\end{array}\sum_{i=u,d,s}(f_{i}+f_{\overline{i}})]\end{array}}\end{array}$........ ( 15a )\\
Using standard techniques {[} 5,6,8,11 ] we can obtain the following
result for the mean free path $\begin{array}{c}
\lambda_{f}\end{array}$:\\
$\begin{array}{c}
{\displaystyle \lambda_{f}^{-1}}=n_{g}\int dq_{\bot}^{2}\frac{{\displaystyle d\sigma_{el}^{gg}}}{{\displaystyle d}q_{\bot}^{2}}=n_{g}\int_{0}^{s/4}dq_{\bot}^{2}{\displaystyle \frac{9}{4}\frac{2\pi\alpha_{s}^{2}}{(q_{\bot}^{2}+m_{D}^{2})^{2}}}=\frac{{\displaystyle {\displaystyle 9n_{g}\pi\alpha_{s}^{2}}}}{{\displaystyle 2m_{D}^{2}(1+\frac{2}{9}\frac{m_{D}^{2}}{T^{2}})}}\end{array}$........ ( 16)\\
which for zero chemical potential using $\begin{array}{c}
{\displaystyle m_{D}^{2}=}\end{array}4\pi\alpha_{s}T^{2}\lambda_{g}$........ ( 17 ) \\
reduces to the well known {[} 8,11 ] result:\\
$\begin{array}{c}
{\displaystyle {\displaystyle \lambda_{f}^{-1}}=\frac{9}{8}a_{1}\alpha_{s}T\frac{1}{1+8\pi\alpha_{s}\lambda_{g}/9}}\end{array}$........ ( 18 )\\
Using standard methods {[} 5,6,8,11 ] we get the modified differential
cross section as\\
$\begin{array}{c}
{\displaystyle \frac{d\sigma_{3}}{{\displaystyle d}q_{\bot}^{2}dy{\displaystyle d^{2}k_{\perp}}}=\frac{{\displaystyle d\sigma_{el}^{gg}}}{{\displaystyle d}q_{\bot}^{2}}}\end{array}{\displaystyle \frac{dn_{g}}{dy{\displaystyle d^{2}k_{\perp}}}}\theta(\lambda_{f}-\tau_{QCD})\theta(\sqrt{s}-k_{\perp}coshy)$........ ( 19 ) 

Recalling the definition of $\begin{array}{c}
{\displaystyle R_{3}=\frac{1}{2}\sigma_{3}n_{g}}\end{array}$........ ( 20 ) with \\
$\begin{array}{c}
{\displaystyle \sigma_{3}=<\sigma(gg\rightarrow ggg)v>}\end{array}$........ ( 21 ), the thermally averaged velocity weighted cross section,
we obtain the required rate as \\
$\begin{array}{c}
{\displaystyle \frac{R_{3}}{T}=\frac{27\alpha_{s}^{3}}{2}\lambda_{f}^{2}n_{g}I(\lambda_{g})}\end{array}$........ ( 22 )\\
where\\
$\begin{array}{c}
{\displaystyle I(\lambda_{g})=\int_{1}^{\sqrt{s}\lambda_{f}}dx\int_{0}^{{\displaystyle \frac{s}{4m_{D}^{2}}}}dz\frac{z}{(1+z)^{2}}[\frac{cosh^{-1}\sqrt{x}}{x\sqrt{[x+(1+z)x_{D}]^{2}-4xzx_{D}}}+}\end{array}$\\
$\begin{array}{c}
{\displaystyle \qquad\qquad\qquad\qquad\qquad\qquad\frac{1}{s\lambda_{f}^{2}}}{\displaystyle \frac{cosh^{-1}\sqrt{x}}{\sqrt{[1+x(1+z)y_{D}]^{2}-4xzy_{D}}}]}\end{array}$........ ( 23 )\\
with\\
$\begin{array}{c}
{\displaystyle x_{D}=m_{D}^{2}\lambda_{f}}\\
{\displaystyle y_{D}=\frac{m_{D}^{2}}{s}}\end{array}$........ ( 24 )

\subsection{{\normalsize Quark Anti-Quark pair production rate}\protect \\
{\normalsize }\protect \\
}

We have parton production rates in the RHS of the number density evolution
equations. Let us evaluate the quark-antiquark pair production reaction
rate $\begin{array}{c}
R_{2q}\end{array}$. We have {[} 4 ]\\
$\begin{array}{c}
R_{2q}=R_{{\displaystyle _{gain}}}^{{\displaystyle gg}}-R_{{\displaystyle _{loss}}}^{{\displaystyle gg}}\end{array}$........ ( 25 )\\
where\\
$\begin{array}{c}
{\displaystyle R_{{\displaystyle _{gain}}}^{{\displaystyle gg}}=\int\frac{d^{3}p_{1}}{(2\pi)^{^{{\displaystyle 3}}}2E_{1}}\int\begin{array}{c}
{\displaystyle \frac{d^{3}p_{2}}{(2\pi)^{^{{\displaystyle 3}}}2E_{2}}}\end{array}\int\begin{array}{c}
{\displaystyle \frac{d^{3}p_{3}}{(2\pi)^{^{{\displaystyle 3}}}2E_{3}}}\end{array}\int\begin{array}{c}
\begin{array}{c}
{\displaystyle \frac{d^{3}p_{4}}{(2\pi)^{^{{\displaystyle 3}}}2E_{4}}(2\pi)^{4}}\end{array}\end{array}}\end{array}$\\
$\begin{array}{c}
{\displaystyle \delta^{4}(p_{1}+p_{2}-p_{3}-p_{4})\Sigma|M_{gg\rightarrow i\overline{i}}|^{{\displaystyle ^{2}}}f_{g}(p_{1})f_{g}(p_{2})(1-f_{q}(p_{3}))(1-f_{\bar{q}}(p_{4}))}\end{array}$... ( 25a )\\
and\\
$\begin{array}{c}
{\displaystyle R_{{\displaystyle _{loss}}}^{{\displaystyle gg}}=\int\frac{d^{3}p_{1}}{(2\pi)^{^{{\displaystyle 3}}}2E_{1}}\int\begin{array}{c}
{\displaystyle \frac{d^{3}p_{2}}{(2\pi)^{^{{\displaystyle 3}}}2E_{2}}}\end{array}\int\begin{array}{c}
{\displaystyle \frac{d^{3}p_{3}}{(2\pi)^{^{{\displaystyle 3}}}2E_{3}}}\end{array}\int\begin{array}{c}
\begin{array}{c}
{\displaystyle \frac{d^{3}p_{4}}{(2\pi)^{^{{\displaystyle 3}}}2E_{4}}(2\pi)^{4}}\end{array}\end{array}}\end{array}$\\
$\begin{array}{c}
{\displaystyle {\displaystyle \delta^{4}(p_{1}+p_{2}-p_{3}-p_{4})\Sigma|M_{gg\rightarrow i\overline{i}}|^{{\displaystyle ^{2}}}(1+f_{g}(p_{1}))(1+f_{g}(p_{2}))f_{q}(p_{3})f_{\bar{q}}(p_{4})}}\end{array}$... ( 25b )\\
Following {[} 4 ], we can say that there are three topologically distinct
Feynmann diagrams that contribute towards the quark-antiquark pair
production process. Evaluating them performing traces and finally
adding them up we can find the net squared matrix element. We basically
follow the lines of {[} 4 ]. Transforming variables as \\
$\begin{array}{c}
{\displaystyle q=p_{1}+p_{2}}\\
{\displaystyle {\displaystyle p=\frac{1}{2}(p_{1}-p_{2})}}\\
{\displaystyle {\displaystyle q'=p_{3}+p_{4}}}\\
{\displaystyle p'=\frac{1}{2}(p_{3}-p_{4})}\end{array}$........ ( 26 )\\
with restrictions\\
$\begin{array}{c}
{\displaystyle q_{0}>2m_{i}}\\
{\displaystyle s=q_{0}^{2}-|\overrightarrow{\mathbf{q|}}}^{2}\geq4m_{i}^{2}\\
{\displaystyle {\displaystyle p_{0}^{2}\leq\frac{q^{2}}{4}}}\\
{\displaystyle p'_{0}.p'_{0}\leq\frac{q^{2}}{4}(1-\frac{4m_{i}^{2}}{s})}\end{array}$........ ( 27 )\\
and transforming the three dimensional integrals to four dimensional
integrals using\\
${\displaystyle \begin{array}{c}
{\displaystyle \int}\end{array}\frac{d^{3}p_{i}}{2E_{i}}=\int d^{4}p_{i}\delta(p^{2}-m_{i}^{2})}$........ ( 28)\\
with the new set of variables \\
$\begin{array}{c}
{\displaystyle q_{0}=-Tln}\mathit{v}+2m_{i}\\
{\displaystyle {\displaystyle {\displaystyle q^{\frac{1}{2}}}=(q_{0}^{2}-4m_{i}^{2})^{\frac{1}{2}}u}}\\
{\displaystyle p_{0}=\frac{q}{2}(1-\frac{4m_{i}^{2}}{s})^{\frac{1}{2}}x}\\
{\displaystyle p'_{0}=\frac{q}{2}y}\end{array}$........ ( 29)\\
we arrive at the rate \\
$\begin{array}{c}
{\displaystyle R_{2g}=\frac{\alpha_{s}^{2}}{2\pi^{3}}T\int_{0}^{1}du\int_{0}^{1}dv\int_{0}^{1}dx\int_{0}^{1}dy\frac{u^{2}}{v}(1-\frac{4m_{i}^{2}}{s})^{\frac{1}{2}}(q_{0}^{2}-4m_{i}^{2})^{\frac{3}{2}}f_{Quarks}f_{phase1}}\end{array}$........ (30 )\\
where\\
$\begin{array}{c}
{\displaystyle f_{Quarks}=f_{g}(\frac{q_{0}}{2}+p_{0})f_{g}(\frac{q_{0}}{2}-p_{0})(1-f_{q}(\frac{q_{0}}{2}+p'_{0}))(1-f_{\overline{q}}(\frac{q_{0}}{2}-p'_{0}))-}\end{array}$\\
$\begin{array}{c}
{\displaystyle \qquad\qquad}\end{array}(1+f_{g}(\frac{q_{0}}{2}+p_{0}))(1+f_{g}(\frac{q_{0}}{2}-p_{0}))f_{q}(\frac{q_{0}}{2}+p'_{0})f_{\overline{q}}(\frac{q_{0}}{2}-p'_{0})$........ ( 30a )\\
and\\
$\begin{array}{c}
{\displaystyle f_{phase1}=A+B[\frac{1}{K_{+}}+\frac{1}{K_{-}}]+C[\frac{\bigtriangleup_{+}}{K_{+}^{3}}+\frac{\bigtriangleup_{-}}{K_{+}^{3}}]}\end{array}$........ ( 30b )\\
with\\
\\
$\begin{array}{c}
{\displaystyle {\displaystyle A=3[1-}[1-{\displaystyle \frac{4m_{i}^{2}}{s}}][{\displaystyle \frac{(1-x^{2})(1-y^{2})}{2}}+x^{2}y^{2}]]-\frac{{\displaystyle 34}}{{\displaystyle 3}}-24{\displaystyle \frac{m_{i}^{2}}{s}}}\end{array}$........ ( 30c )\\
$\begin{array}{c}
{\displaystyle {\displaystyle B=\frac{16}{3}[1+\frac{4m_{i}^{2}}{s}+\frac{m_{i}^{4}}{s^{2}}]}}\end{array}$........ ( 30d )\\
$\begin{array}{c}
{\displaystyle {\displaystyle C=-\frac{128}{3}{\displaystyle \frac{m_{i}^{4}}{s^{2}}}}}\end{array}$........ ( 30e )\\
$\begin{array}{c}
{\displaystyle K_{\pm}={\displaystyle [1-}[1-{\displaystyle \frac{4m_{i}^{2}}{s}}][(1-x^{2}-y^{2})\pm2[1-{\displaystyle \frac{4m_{i}^{2}}{s}}]^{\frac{1}{2}}xy]^{\frac{1}{2}}}\end{array}$........ ( 30f )\\
$\begin{array}{c}
{\displaystyle \bigtriangleup_{\pm}=1\pm}\end{array}[1-{\displaystyle \frac{4m_{i}^{2}}{s}}]^{\frac{1}{2}}xy$........ ( 30g )\\

\subsection{{\normalsize Quark Flavour Changing Rate}\protect \\
}

For the quark flavour changing process we have {[} 4 ]\\
$\begin{array}{c}
R_{qg}={\displaystyle R_{gain}^{{\displaystyle q\overline{q}}}}-R_{{\displaystyle _{loss}}}^{{\displaystyle q\overline{q}}}\end{array}$........ ( 31 )\\
where\\
$\begin{array}{c}
{\displaystyle R_{{\displaystyle _{gain}}}^{q\overline{q}}=\int\frac{d^{3}p_{1}}{(2\pi)^{^{{\displaystyle 3}}}2E_{1}}\int\begin{array}{c}
{\displaystyle \frac{d^{3}p_{2}}{(2\pi)^{^{{\displaystyle 3}}}2E_{2}}}\end{array}\int\begin{array}{c}
{\displaystyle \frac{d^{3}p_{3}}{(2\pi)^{^{{\displaystyle 3}}}2E_{3}}}\end{array}\int\begin{array}{c}
\begin{array}{c}
{\displaystyle \frac{d^{3}p_{4}}{(2\pi)^{^{{\displaystyle 3}}}2E_{4}}(2\pi)^{4}}\end{array}\end{array}}\end{array}$\\
$\begin{array}{c}
{\displaystyle \delta^{4}(p_{1}+p_{2}-p_{3}-p_{4})\Sigma|M_{s\overline{s}\rightarrow q\overline{q}}|^{{\displaystyle ^{2}}}}\end{array}f_{q}(p_{1})f_{\overline{q}}(p_{2})(1-f_{s}(p_{3}))(1-f_{\bar{s}}(p_{4}))$... ( 31a )\\
and\\
$\begin{array}{c}
{\displaystyle R_{{\displaystyle _{loss}}}^{q\overline{q}}=\int\frac{d^{3}p_{1}}{(2\pi)^{^{{\displaystyle 3}}}2E_{1}}\int\begin{array}{c}
{\displaystyle \frac{d^{3}p_{2}}{(2\pi)^{^{{\displaystyle 3}}}2E_{2}}}\end{array}\int\begin{array}{c}
{\displaystyle \frac{d^{3}p_{3}}{(2\pi)^{^{{\displaystyle 3}}}2E_{3}}}\end{array}\int\begin{array}{c}
\begin{array}{c}
{\displaystyle \frac{d^{3}p_{4}}{(2\pi)^{^{{\displaystyle 3}}}2E_{4}}(2\pi)^{4}}\end{array}\end{array}}\end{array}$\\
$\begin{array}{c}
{\displaystyle {\displaystyle \delta^{4}(p_{1}+p_{2}-p_{3}-p_{4})\Sigma|M_{s\overline{s}\rightarrow q\overline{q}}|^{{\displaystyle ^{2}}}(1-f_{q}(p_{1}))(1-f_{\overline{q}}(p_{2}))f_{s}(p_{3})f_{\bar{s}}(p_{4})}}\end{array}$... ( 31b )\\
Following {[} 4 ], we can say that there is only one type of topologically
distinct Feynmann diagram that contributes towards the quark flavour
changing process or the strange quark pair production process. Evaluating
it, performing trace calculations we can find the squared matrix element.
We basically follow the lines of {[} 4 ]. Transforming variables as
in the case before and performing identical operations we can get
the rate as \\
$\begin{array}{c}
{\displaystyle \begin{array}{c}
{\displaystyle R_{qg}=\frac{\alpha_{s}^{2}}{2\pi^{3}}T\int_{0}^{1}du\int_{0}^{1}dv\int_{0}^{1}dx\int_{0}^{1}dy\frac{u^{2}}{v}(1-\frac{4m_{i}^{2}}{s})^{\frac{1}{2}}(q_{0}^{2}-4m_{i}^{2})^{\frac{3}{2}}f_{Strange}f_{phase2}}\end{array}}\end{array}$........ ( 32 )\\
with\\
$\begin{array}{c}
{\displaystyle f_{Quarks}=f_{q}(\frac{q_{0}}{2}+p_{0})f_{\overline{q}}(\frac{q_{0}}{2}-p_{0})(1-f_{s}(\frac{q_{0}}{2}+p'_{0}))(1-f_{s}(\frac{q_{0}}{2}-p'_{0}))-}\end{array}$\\
$\begin{array}{c}
{\displaystyle \begin{array}{c}
{\displaystyle \qquad\qquad}\end{array}(1+f_{s}(\frac{q_{0}}{2}+p_{0}))(1+f_{s}(\frac{q_{0}}{2}-p_{0}))f_{q}(\frac{q_{0}}{2}+p'_{0})f_{\overline{q}}(\frac{q_{0}}{2}-p'_{0})}\end{array}$... ( 32a )\\
$\begin{array}{c}
{\displaystyle f_{phase2}={\displaystyle [1+}[1-{\displaystyle \frac{4m_{i}^{2}}{s}}][{\displaystyle \frac{(1-x^{2})(1-y^{2})}{2}}+x^{2}y^{2}]]+\frac{4m_{i}^{2}}{s}}\end{array}$...( 32b )\\

\section{Results}

\subsection{Initial Conditions\protect \\
}

As an initial condition at the point of thermalization ( i.e the point
after which the system evolves according to the laws of hydrodynamics)
we have predictions from the HIJING and SSPC models {[} 15 ]. In addition
to these inputs we shall require inputs for at least two of the following:
initial baryon number density, initial light quark chemical potential
and ratio of initial non-equilibrium fugacities of light quarks and
antiquarks. 

In absence of initial values thereof ( although we have some initial
conditions calculated at $\begin{array}{c}
\tau_{i}=0.1\end{array}fm/c$ {[}1,16 ], which is at a much earlier time than the time of thermalization
at $\begin{array}{c}
\begin{array}{c}
\tau_{i}=0.25\end{array}fm/c\end{array}$ for SSPC initial conditions and earlier still for HIJING at $\begin{array}{c}
\begin{array}{c}
\begin{array}{c}
\tau_{i}=0.6-0.7\end{array}fm/c\end{array}\end{array}$ {[} 15 ] ) let us study the relative degree of equilibration for
varying initial conditions although it must be emphasized that most
of these initial conditions would be of purely academic interest only,
as they cannot be realised in practice at the collider experiments.
Nevertheless, it is hoped that this exercise would pave the way for
a clearer understanding of the physical processes involved. 

Even though two inputs remain rather arbitrary, we use both set of
initial conditions ( LHC and RHIC ) to study the trends. For the usual
inputs we use the following sets of data ( SSPC )\\
\\
\textbf{LHC}

\begin{tabular}{|c|c|c|c|c|}
\hline 
$\begin{array}{c}
\tau_{i}(fm/c)\end{array}$&
$T(GeV)$&
$\begin{array}{c}
\lambda_{g}\end{array}$&
$\begin{array}{c}
\lambda_{q}\end{array}$&
$\begin{array}{c}
\lambda_{s}\end{array}$\tabularnewline
\hline
\hline 
0.25&
1.02&
0.43&
0.086&
0.043\tabularnewline
\hline
\end{tabular}\\
\\
\textbf{RHIC}

\begin{tabular}{|c|c|c|c|c|}
\hline 
$\begin{array}{c}
\begin{array}{c}
\tau_{i}(fm/c)\end{array}\end{array}$&
$\begin{array}{c}
T(GeV)\end{array}$&
$\begin{array}{c}
\begin{array}{c}
\lambda_{g}\end{array}\end{array}$&
$\begin{array}{c}
\begin{array}{c}
\lambda_{q}\end{array}\end{array}$&
$\begin{array}{c}
\begin{array}{c}
\lambda_{s}\end{array}\end{array}$\tabularnewline
\hline
\hline 
0.25&
0.668&
0.34&
0.064&
0.032\tabularnewline
\hline
\end{tabular}

As a representative plot we use baryon density 0.15/fm\textasciicircum{}3
and ratio=1.5 with the initial conditions as given above. We also
try to see the effect of including the quark flavour changing processes
( qfcp).

\subsection{Observed trends\protect \\
}

We study comparatively the outputs with given initial baryon number
density and initial light quark to anti-quark fugacity ratio. We vary
the baryon number density between 0.21/fm\textasciicircum{}3 and 0.11/fm\textasciicircum{}3
while we change the ratio between 1.9 and 1.1. We arrive at the following
conclusions :\\

1. The nature of variations of the physical quantities are more or
less in the line of earlier works. The results for RHIC and LHC initial
conditions are shown in figures 1 and 2. \textcolor{black}{Temperature,Non
equilibrium fugacity and Chemical Potential variations with RHIC initial
conditions are given in figure 1 while that for LHC are given in figure
2. The decaying curves give temperature variations while the positive
rising curves are sequentially (from top) for gluon, light quark,
light anti-quark and strange quark fugacity variations. Plots for
both quark flavour changing process included and excluded cases are
shown. For identification please see point 4 below.}

We observe that \textcolor{red}{\large }\\
\textcolor{black}{i) As expected, the temperature falls with time
while the non-equilibrium fugacities increase.}\\
\textcolor{black}{ii) Contrary to the Juttner case, the chemical potential
remains negative all along and as expected approaches zero as the
system equilibrates.}\\
\textcolor{black}{iii) The QGP,as expected remains to be gluon dominated.}\\
\textcolor{black}{iv) For a system at higher chemical potential, the
temperature falls at a slower rate signifying lesser amount of energy
expenditure to create partons which shows up in the slower rise of
all partons except for the light anti-quark, which shows a larger
growth rate due to the presence of the exponentiated chemical potential.}\textcolor{blue}{\large }\\
\\

2. For a given initial ratio, except for the light quarks and anti-quarks,
the output does not depend much on the initial baryon number density.
For the light quarks and antiquarks this variation is due to the presence
of the respective non-zero chemical potentials. \\

3. For a given initial baryon number density we can recast the equation
for baryon number density in the form \[
e^{\mu_{q}/T}=\frac{C}{2\lambda_{q}}+\sqrt{\frac{1}{D}+\frac{C^{2}}{4\lambda_{q}^{2}}}\]
$\qquad\qquad\qquad\qquad\qquad\qquad\qquad\qquad\qquad\qquad\qquad$........
( 33 )

where C is a constant for a given temperature, light quark fugacity
and given baryon number density. Here D is the light quark to antiquark
initial fugacity ratio. For a fixed light quark fugacity and temperature,
as D falls clearly the RHS of the above equation increases which indicates
a rise in the chemical potential. Again for a system of higher chemical
potential, the temperature has to drop at a slower rate due to the
constraint imposed. Hence we observe that\\
\\
i) The temperature falls at a slower rate for a smaller value of the
light quark to antiquark initial fugacity ratio.\\
ii) As the temperature falls at a slower rate, it would imply a lesser
expenditure of energy to produce particles in general . This would
show up in the slower rise of all fugacity values except the light
antiquark, which would show a higher growth rate. This is due to the
exponentiated chemical potential part. \\

4. For inclusion of the quark-flavour-changing process we observe
the following:\\
\\
 Due to the additional production of s-quarks from massless quarks
via qfcp, we see an additional increase in fugacity of the strange
quark while, the rate of equilibration falls for all other non-strange
quarks . \\

5. For the chemical potential of the light quarks, clearly the value
thereof tends to level off  as the system equilibrates . The chemical
potential remains negative all along, which can be attributed to the
choice of the distribution function.\\

\section{Acknowledgement\protect \\
}

\textcolor{black}{The author gratefully acknowledges helpful discussions
with Prof. Bikash Sinha and Prof.Binayak Dutta Roy as also valuable
e-mail clarifications and encouragement from Profs. Z.J.He and Y.G.Ma
during the initial stages of the present work.}

\section{References}

1. N.Hammon, H.Stocker and W.Greiner, Phys. Rev. C61(1999) 014901
\\
2. STAR Collaboration, Phys. Lett. B567(2003) 167 \\
3. BRAHMS Collaboration, Phys. Lett. B607 (2005) 42\\
4. T.Matsui, B.Svetitsky and L.D.McLerran, Phys. Rev. D34(1986) 783
\\
5. D.Dutta, A.K.Mohanty, K.Kumar and R.K.Choudhury, Phys. Rev.C60
(1999)014905\\
6. D.Dutta, A.K.Mohanty, K.Kumar and R.K.Choudhury, Phys. Rev.C61
(2000)064911\\
7. Z.J.He, J.L.Long, Y.G.Ma, G.L.Ma and B.Liu, Phys. Rev. C69(2004)
034906 \\
8. Dipali Pal,Abhijit Sen,Munshi Golam Mustafa and Dinesh Kumar Srivastava,Phys.
Rev. C65(2002)034901 \\
 9. T.S.Biro, E van Doorn, B.Muller, M.H.Thoma and X.N.Wang, Phys.
Rev. C48, 1275 (1993)\\
10. Peter Levai and X.-N.Wang, Proceedings of Strangeness '95,Tucson,Arizona,USA,Jan
4-6,1995,Am.Inst.of Phys.\\
11. Introduction to High Energy Heavy Ion Collisions, C.Y.Wong, World
Scientific, 1994, Chapter-9\\
12. Li Xiong and Edward V Shuryak, Phys. Rev. C49 (1994) 2203\\
13. F.A. Berends,R.Kliess,P.De Causmaecker,R.Gastmans and Tai Tsun
Wu,Phys. Lett. 103B, 124 ( 1981)\\
14. Fred Cooper, Chung Wen Kao and Gouranga C. Nayak, arXiv.org/hep-ph/0207370
and reference therein\\
 15. Munshi G Mustafa and Markus H Thoma, Phys. Rev. C62,(2000)014902\\
 16. E.J.Eskola,Prog.Theo. Phys.Suppl.,129(1997)1-10\\

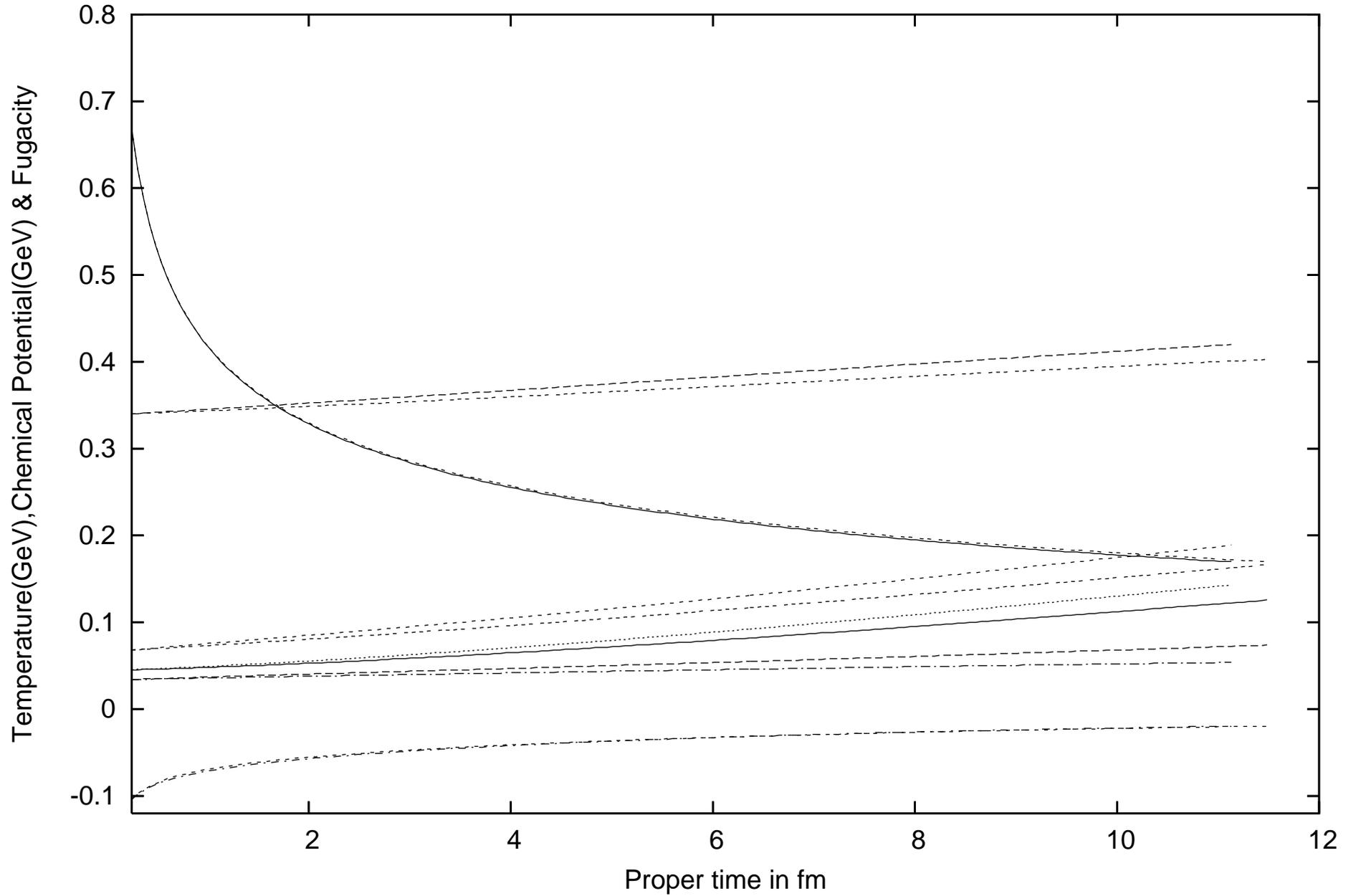
\end{document}